\documentclass[aps,prl,twocolumn,showpacs,groupedaddress]{revtex4}
\usepackage{graphicx}

\begin{document}

\title{Spin waves in a Bose-condensed atomic spin chain}

\author{Weiping Zhang, Han Pu, Chris Search and Pierre
Meystre}
\affiliation{Optical Sciences Center, The University of
Arizona, Tucson, AZ 85721}
\date{\today}

\begin{abstract}
The spin dynamics of atomic Bose-Einstein condensates confined in
a one-dimensional optical lattice is studied. The condensates at
each lattice site behave like spin magnets that can interact with
each other through both the light-induced dipole-dipole
interaction and the static magnetic dipole-dipole interaction. We
show how these site-to-site dipolar interactions can distort the
ground state spin orientations and lead to the excitation of spin
waves. The dispersion relation of the spin waves is studied and
possible detection schemes are proposed.
\end{abstract}
\pacs{PACS numbers: 03.75.Fi, 75.45.+j, 75.60.Ej}
\maketitle

Rapid experimental progress in the realization of trapped
degenerate quantum atomic gases has generated fascinating
opportunities to study a wide range of physical phenomena in
atomic physics, condensed matter physics and quantum optics. In
particular, the recent success in all-optical confinement
\cite{yale1,hansch} and formation \cite{chapman} of atomic
Bose-Einstein condensates provides a unique tool to explore the
magnetic properties and spin-dependent dynamics of ultracold
atomic gases \cite{Meystre}. In this letter, we propose a scheme
to study the excitation and propagation of spin waves in an array
of atomic spinor Bose-Einstein condensates confined or created in
an optical lattice.

\begin{figure}
\includegraphics*[width=7cm,height=3cm]{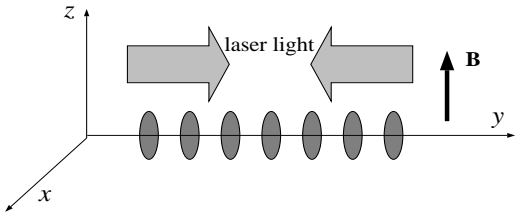}
\caption{Schematic diagram shows that a spinor condensate is
confined in an optical lattice in the $y$-axis with an external
magnetic field along the $z$ direction.} \label{fig1}
\end{figure}

Spin-wave phenomena play an important role in solid state physics
\cite{solid}. In solids, spin-wave excitations result from the
exchange interaction of electrons between atoms in the crystal.
They are usually associated with spin-1/2 Fermi systems with an
effective interaction range of a few angstroms, a typical lattice
period in solid materials. Although the Bose-condensed atoms in
optical lattices exhibit a number of close analogies to atoms in a
real crystal, a number of differences exist. (1) The atomic
spacing in an optical lattice is of the order of half an optical
wavelength and as such is much larger than a crystal lattice
period. (2) As a result, the electron exchange interaction is
completely negligible: spin waves, if they exist, must be caused
by other forms of long range interactions. (3) There is a large
number of (bosonic) atoms at each lattice site, typically of the
order of 1000 or more, and they are subject to Bose-enhancement
effects. As a result of these differences, the Bose condensates in
an optical lattice offer a totally new environment to study spin
dynamics in periodic structures.

The schematic diagram of Fig.~\ref{fig1} shows the system
discussed in this letter. We consider for concreteness a
one-dimensional (1D) optical lattice formed by two $\pi$-polarized
laser beams counter-propagating along the $y$-axis. We assume
that, in the $x$-$z$ plane, Bose-condensed alkali atoms in their
hyperfine ground state manifold are tightly confined by an optical
dipole potential arising from either the transverse profile of the
lattice field or from a separate laser. Hence a 1D coherent atomic
chain is formed along the $y$-axis. We employ a spinor atomic
field theory to describe the interaction of the atoms with the
lattice laser beams. For large detunings
$\Delta=\omega_L-\omega_a$ between the frequency $\omega_L$ of the
laser fields and the atomic transition frequency $\omega_a$ it is
possible to adiabatically eliminate the excited atomic state field
operator. The resulting Heisenberg equations of motion for the
hyperfine ground-state atomic field operators $\hat{\psi}_m({\bf
r},t)$, including interatomic collisions and static magnetic
dipole-dipole interaction, take the form
\cite{Zhang1,Zhang2,ho,ohmi,law},

\begin{eqnarray}
i \hbar \frac{\partial \hat{\psi}_m}{\partial
t}=\left[-\frac{\hbar^2
\nabla^2}{2m} + V_L({\bf r})\right]\hat{\psi}_m + \nonumber \\
\sum_{m',n',n} \int d{\bf r}\, [Q_{mm'n'n}({\bf r},{\bf r'})+
V^{{\rm coll}}_{mm'n'n}({\bf r}-{\bf r'}) \nonumber \\ + V^{{\rm
dd}}_{mm'n'n}({\bf r}-{\bf r'})]\hat{\psi}_{m'}^{\dagger}({\bf
r'})\hat{\psi}_{n'}({\bf r'})\hat{\psi}_n({\bf r})\,, \label{eq1}
\end{eqnarray}
where $V_L({\bf r})=U_0 \exp[-r_{\bot}^2/W_L^2] \cos^2(k_L y)$ is
the light-induced lattice potential, with $r_{\bot} \equiv
\sqrt{x^2+z^2}$. The potential depth is defined as $U_0=\hbar
|\Omega|^2 /6\Delta$ with $\Omega$ being the Rabi frequency,
$k_L=2 \pi /\lambda_L$ the wave number of the lattice beams, and
$W_L$ the beam width. The index $m=-F,...F$ denotes the Zeeman
sublevels of the electronic ground state of the atoms with angular
momentum $F$. In this letter we take $F=1$ for the ground state
alkali atoms.

The first nonlinear term in Eq.~(\ref{eq1}), originates from the
photon-exchange interaction between the condensed atoms. It
describes the light-induced dipole-dipole interaction and is
characterized by the quantity
\begin{widetext}
\begin{equation}
Q_{mm'n'n}({\bf r},{\bf r'})= 3 \frac{\gamma}{\Delta}U_0 \exp
\left(-\frac{r_{\bot}^2+{r'}_{\bot}^{2}}{W_L^2} \right) \cos(k_L
y) \cos(k_L y')\left[ \frac{4}{9} \delta_{m'n'} \delta_{mn} {\bf
e_0} \cdot {\bf W}({\bf r}-{\bf r'})\cdot {\bf e_0}-{\bf d_{m'n'}}
\cdot {\bf W}({\bf r}-{\bf r'}) \cdot {\bf d_{mn}} \right],
\label{eq2}
\end{equation}
\end{widetext}
where $\gamma$ is the single-atom spontaneous emission rate and
${\bf d_{mn}}=(F_{mn}^{(+)} {\bf e_{-1}} + F_{mn}^{(-)}{\bf
e_{+1}})/6 \hbar$ denotes the dipole moments induced by the
$\pi$-polarized light fields. Here, $F_{mn}^{(\pm)}$ are the
matrix elements of the ``+'' and ``$-$'' components of the total
angular momentum operator ${\bf F}$, and ${\bf e_{\pm 1,0}}$ are
unit vectors in the spherical harmonic basis. The tensor ${\bf
W}$, describing the spatial profile of the light-induced
dipole-dipole interaction, has the form
\begin{eqnarray}
{\bf W}({\bf {r}})=\frac{3}{4}\left[({\bf 11}-3\hat{{\bf
r}}\hat{{\bf r}}
)\left(\frac{\sin\xi}{\xi^2}+\frac{\cos\xi}{\xi^3}\right) -({\bf
11}-\hat{{\bf r}}\hat{{\bf r}})\frac{\cos\xi}{\xi} \right],
\label{eq3}
\end{eqnarray}
where ${\bf 11}$ is the unit tensor, $\hat{{\bf r}}={\bf r}/|{\bf
r}|$ and $\xi =k_L|{\bf r}|$.

The two-body ground-state collisions and magnetic dipole-dipole
interaction are described by the potentials
\begin{eqnarray*}
V^{{\rm coll}}_{mm'n'n}({\bf r}-{\bf r'}) &=& \left[\lambda_s
\delta_{m'n'} \delta_{mn} + \lambda_a {\bf F}_{mn} \cdot {\bf
F}_{m'n'} \right] \delta({\bf r}-{\bf r'}), \\
V^{{\rm dd}}_{mm'n'n}({\bf r}-{\bf r'}) &=& \frac{\mu_0 \mu_B^2
g_F^2}{4
\pi |{\bf r}-{\bf r'}|^3} [{\bf F}_{mn} \cdot {\bf F}_{m'n'} \nonumber \\
&&-3\frac{{\bf F}_{mn} \cdot ({\bf r}-{\bf r'}) {\bf F}_{m'n'}
\cdot ({\bf r}-{\bf r'})}{|{\bf r}-{\bf r'}|^2}],
\end{eqnarray*}
respectively, where $\lambda_s$ and $\lambda_a$ are related to the
$s$-wave scattering lengths of the spinor condensate\cite{law}.

The optical potential associated with a sufficiently deep optical
lattice is equivalent to a periodic array of independent
``microtraps''\cite{hansch}. If the depth of each of those is
large enough, an array of independent Bose condensates can be
formed in the lattice, and it is convenient to expand the spinor
atomic field operator as
\begin{eqnarray}
\hat{\psi}_m({\bf r})= \sum_{i} \phi_i({\bf r})\hat{a}_m(i)\,,
\label{eq4}
\end{eqnarray}
where $\phi_i$ is the condensate wave function for the $i$th
microtrap and the operators $\hat{a}_m(i)$ satisfy the bosonic
commutation relations $[\hat{a}_m(i), \hat{a}^\dag_n(j)] =
\delta_{mn}\delta_{ij}$. For deep enough microtraps the spatial
overlap between the individual condensate wave functions is
negligible, and they can be considered as independent. Under this
tight-binding condition, the spatial wave function of the $i$th
condensate is then determined by the Gross-Pitaevskii equation
$\left[-\frac{\hbar^2 \nabla^2}{2m} + V_i({\bf r}) + \lambda_s
(N_i-1)|\phi_i({\bf r})|^2 \right]\phi_i({\bf r})=\mu_i
\phi_i({\bf r})$, with $V_i({\bf r})$ being the potential near the
$i$-th microtrap and $N_i=\sum_{m} \langle \hat{a}^{\dagger}_m(i)
\hat{a}_m(i)\rangle$ the number of condensed atoms at the site,
and $\mu_i$ the chemical potential.

For $F=1$, the individual condensates consist of atoms with three
Zeeman sublevels, hence they behave as collective spin magnets in
the presence of external magnetic fields or spin-dependent
interactions. Such spin magnets, localized along the lattice axis,
form a 1D coherent $Bose$ atomic spin chain. Using Eq.~(\ref{eq1})
and Eq.~(\ref{eq4}), and ignoring both the non-resonant and
spin-independent constant terms, we can construct the Hamiltonian
describing this spin chain,
\begin{eqnarray}
H &=& \sum_{i} [\lambda_a' \hat{{\bf S}}_i^2 - \gamma_B \hat{{\bf
S}}_i \cdot
{\bf B} - \sum_{j \neq i} J^z_{ij} \hat{S}^z_i \hat{S}^z_j  \nonumber \\
&& - \sum_{j \neq i} J_{ij}(\hat{S}^{(-)}_i \hat{S}^{(+)}_j +
\hat{S}^{(+)}_i \hat{S}^{(-)}_j)], \label{eq5}
\end{eqnarray}
where we have defined the collective spin operators $\hat{{\bf
S}}_i = \sum_{mn}\hat{a}^{\dagger}_m(i) {\bf F}_{mn}
\hat{a}_n(i)$, with components  $\hat S_i^{\{\pm, z \}}$. We have
also introduced an external magnetic field ${\bf B}=B_0 {\bf e_0}$
whose strength is strong enough to polarize the ground state spin
orientations of the atomic chain along the quantization axis
$z$\cite{Meystre}. The parameter $\gamma_B = -\mu_B g_F$ is the
gyromagnetic ratio, $\mu_B$ being the Bohr magneton and $g_F$ the
Land{\' e} $g$-factor.

The first term in Hamiltonian (\ref{eq5}) results from the
spin-dependent interatomic collisions at a given site, with
$\lambda_a' = (1/2) \lambda_a \int d^3r |\phi_i({\bf r})|^4$. The
last two terms describe the site-to-site spin coupling induced by
both the static magnetic field and light-induced dipolar
interactions. The coupling coefficients have the explicit forms
\begin{eqnarray}
J^z_{ij} &=& \frac{\mu_0 \gamma_B^2}{16 \pi \hbar^2} \int d{\bf r}
\int d{\bf r'} \frac{|{\bf r'}|^2-3y'^2}{|{\bf r'}|^5}
|\phi_i({\bf r})|^2 |\phi_j({\bf r}-{\bf r'})|^2 ,\nonumber \\
J_{ij} &=& \frac{\gamma U_0}{24 \Delta \hbar^2 k_L^3} \int d{\bf
r} \int d{\bf r'} f_c({\bf r'}) \exp\left(-\frac{r_\bot^2+|{\bf
r}_\bot - {\bf r'}_\bot|^2}{W_L^2} \right)
\nonumber \\
&& \cos(k_L y) \cos[k_L (y-y')] {\bf e}_{+1} \cdot {\bf W}({\bf
r'})
\cdot {\bf e}_{-1} \nonumber \\
&& |\phi_i({\bf r})|^2 |\phi_j({\bf r}-{\bf r'})|^2 +\frac{1}{2}
J^z_{ij} \, ,\label{eq6}
\end{eqnarray}
where we have introduced a cut-off function $f_c({\bf
r})=\exp(-r^2/L_c^2)$ to describe the effective interaction range
of the light-induced dipole-dipole interaction, $L_c = N \gamma/c$
being the coherence length associated with the collective
spontaneous emission of $N$ atoms\cite{cutoff}.

The physics implicit in Hamiltonian (\ref{eq5}) is quite clear.
Consider first the situation without site-to-site coupling,
$J_{ij}=0$. In the presence of a sufficiently strong external
magnetic field, the spins align themselves along the quantization
axis $z$. For ferromagnetic condensates as in the case of
$^{87}$Rb ($\lambda'_a < 0$)\cite{Heinzen,jila},
--- or in the presence of a strong external magnetic field for
anti-ferromagnetic condensates --- the ground state of the
Hamiltonian is $|GS \rangle = |N, -N \rangle$, where $N =\sum_i
N_i$ is the total atomic number in the lattice. The total spin at
site $i$ has the expectation value $\langle \hat{S}^z_i \rangle =
-N_i \hbar$, where the factor $N_i$ is due to Bose enhancement.

For $J_{ij} \neq 0$, the situation changes drastically: the
transfer of transverse spin excitation from site to site is
allowed, resulting in the distortion of the ground state spin
structure. This distortion can propagate from site to site and
hence generate spin waves along the Bose condensed atomic spin
chain. From Hamiltonian (\ref{eq5}), we can derive the Heisenberg
equations of motion for the spin excitations as
\begin{eqnarray}
i \hbar \frac{\partial \hat{S}^{(-)}_q}{\partial t} = (\omega_0 +
\Delta \omega_q) \hat{S}^{(-)}_q -\sum_{j \neq q} \chi_{qj}
\hat{S}^{(-)}_j ,\label{eq7}
\end{eqnarray}
where we have replaced the spin operator $\hat{S}^z_q$ by its
ground state expectation value in the mean-field approximation.
The frequencies $\omega_0 = -\gamma_B B$ and $\Delta \omega_q = 2
\sum_{j \neq q} J^z_{qj} N_j \hbar$ describe the precessing of the
$q$th spin caused by the external magnetic field and the static
magnetic dipolar interaction. The site-to-site spin coupling
coefficients $\chi_{qj}=2 J_{qj} N_q \hbar$ determine the
propagation of the spin waves. From Eq.~(\ref{eq6}), we observe
that the light-induced dipolar interaction only contributes to
spin coupling in the $x$-$y$ plane. This is because the
$\pi$-polarized lattice beams only induce an effective dipole
moment in that plane.

To further understand how Eq.~(\ref{eq7}) determines the existence
and propagation of spin waves, it is helpful to consider the
special case where the lattice is infinitely long and the spin
excitations are in the long-wavelength limit. We can then
reexpress Eq.~(\ref{eq7}) in its continuous limit by the
replacements $\hat{S}^{(-)}_q \rightarrow S(y,t)$, $\chi_{qj}
\rightarrow \chi(y-y')$, and $(\omega_0 + \Delta \omega_q)
\rightarrow \omega(y)$,
\begin{eqnarray}
i \frac{\partial S(y,t)}{\partial t} = \omega(y) S(y,t)
-\frac{2}{\lambda_L} \int dy' \chi(y-y') S(y',t). \label{eq8}
\end{eqnarray}

In the long-wavelength limit, the last integral term in
Eq.~(\ref{eq8}) can be evaluated up to the second-order expansion
of $S(y',t)$. This leads to an effective Schr\"odinger equation
\begin{eqnarray}
i \frac{\partial S(y,t)}{\partial t} = \left[-\frac{\beta_1}{2}
\frac{\partial^{2}}{\partial y^{2}} - \beta_0 + \omega(y)\right]
S(y,t) \label{eq9}
\end{eqnarray}
where we have defined $\beta_{n}=(2/\lambda_L) \int d\eta
\chi(\eta) \eta^{2n}$ for $n=0,1$. Evidently, from
Eq.~(\ref{eq9}), $S(y,t)$ describes the ``waves" caused by spin
excitations in the $x$-$y$ plane. These waves can be associated
with the center-of-mass motion of a quantum-mechanical particle of
effective mass $m=\hbar/\beta_1$. Being similar to the phonon
associated with sound waves, the excitations associated with the
spin waves are usually referred to as ``magnons"\cite{solid}.

\begin{figure}
\includegraphics*[width=8cm,height=5cm]{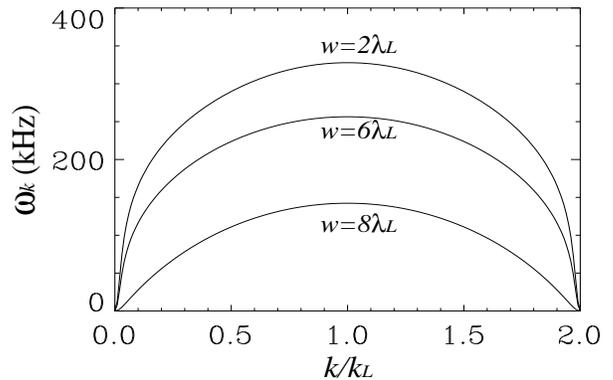}
\caption{Magnon dispersion spectrum. In the calculation, we have
assumed that the spatial dimension along $y$-axis of the
condensate in each lattice site is much less than $\lambda_L$,
while in the transverse ($x$-$z$) plane, the condensate has a
Gaussian shape with a width $w$. We have taken $\lambda_L = 1
\mu$m and $\gamma |\Omega|^2/\Delta^2 =10^3$. We have used a total
number of 100 lattice sites with 2000 atoms in each site. }
\label{fig2}
\end{figure}
We can evaluate the magnon dispersion relation by solving the
discrete wave equation (\ref{eq7}) for waves of the form
$\hat{S}^{(-)}_q = \hat{\alpha_k} \exp[-i(\omega_0 + \Delta
\omega_0 -2 \sum_{j > 0} \chi_{0j})t]
\exp[i(kq\lambda_L/2-\omega_k t)]$. For simplicity, we assume a
lattice with each site having the same number of atoms, $N_i=N_0$.
This yields the magnon dispersion relation
\begin{eqnarray}
\omega_k = 2 \sum_{j > 0} \chi_{0j} [1-\cos(jk\lambda_L /2)] ,
\label{eq10}
\end{eqnarray}
where we have taken $q=0$ in the coefficients of Eq.(\ref{eq7})
since their values are independent of $q$ in the case at hand as
long as we have a sufficient number of lattice sites.

Figure~\ref{fig2} shows the magnon dispersion spectrum and its
dependence on the transverse width $w$ of the Bose condensates. As
$w$ increases the dipolar interactions among atoms tend to cancel
each other, thereby reducing the excitation frequency of the spin
waves. As a result, it becomes difficult to excite the spin waves
in the coherent atomic chain through the dipole-dipole interaction
for condensate widths much larger than one optical wavelength. In
addition, our numerical calculations show that for typical
experimental parameters the strength of the light-induced dipolar
interaction dominates over the magnetic dipolar interaction in the
determination of the coupling coefficients $\chi_{0j}$.

\begin{figure}
\includegraphics*[width=7cm,height=5cm]{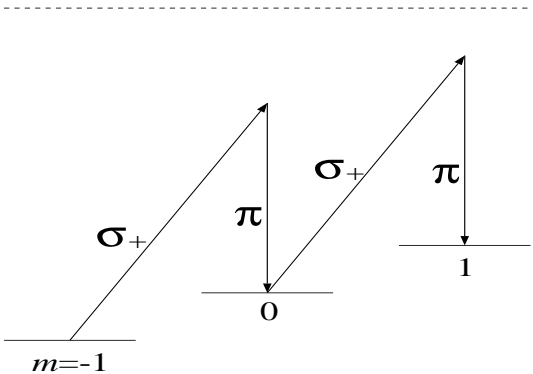}
\caption{Magnon detection schemes.} \label{fig3}
\end{figure}

Having established the existence of spin waves in condensate
lattices, the question remains to determine how to detect them.
Any optical or magnetic method which can excite the internal
transitions between the atomic Zeeman sublevels can be used for
this purpose. On the other hand, the detection scheme should be
able to distinguish different spin wave modes (see discussion
below). A natural choice consists in employing Raman transitions,
as shown in Fig.~\ref{fig3}. In this case the combination of a
circularly-polarized and a $\pi$-polarized Raman beam will couple
all three Zeeman sublevels. This exactly creates the spin
transition associated with the operators $\hat{S}^{(\pm)}$. The
existence of spin waves can then be detected by measuring the
absorption of one of the Raman beams.
The absorption spectrum is proportional to the sum of the
transition probabilities among the Zeeman sublevels,
\begin{eqnarray*}
P \propto \sum_k \frac{|\langle \Phi_k|S_q^{+}|GS\rangle|^2}{2}
\frac{\sin^2[(\nu-\omega_p-\omega_k)\tau/2]}
{[(\nu-\omega_p-\omega_k)\tau/2]^2} \,,
\end{eqnarray*}
where the kets $|\Phi_k \rangle$ denote excited states with spin
waves of frequency $\omega_k$, $\nu$ is the frequency difference
between the two Raman beams, $\tau$ is the measurement time and
$\omega_p = \omega_0 + \Delta \omega_0 -2 \sum_{j>0} \chi_{0j}$
defines the total spin precessing frequency. Absorption resonances
occur whenever the frequency $\nu$ is tuned to match a spin-wave
frequency $\omega_k$. For an infinitely long lattice, this will
produce a broadened absorption spectrum whose width characterizes
the existence of spin waves. In practice, though, the optical
lattice is finite with length $L$. Such lattices only allow the
excitation of spin waves with discrete wave numbers resulting from
the resonance excitation condition $k L=n \pi$, $k$ being the wave
number of the spin waves in Eq.~(\ref{eq10}). As a result, the
absorption spectrum will exhibit a multi-peak structure. In the
long-wavelength limit, the interval between peaks is proportional
to
\[ \Delta \omega_k \approx (2n-1) \frac{\pi^2}{N_L^2}
 \sum_{j>0} \chi_{0j} j^2\,,\]
with $N_L=L /(\lambda_L/2)$ being the total number of lattice
sites.

In current experiments in optical lattices, $N_L$ is in the range
of $10 \sim  100$, and each lattice site can accommodate a few
thousand atoms, in which case $\sum_{j>0} \chi_{0j} j^2$ can reach
values of about a few MHz. This leads to a requirement for the
frequency measurement precision of about $10 \sim 100$KHz. This is
achievable with current techniques.

Alternatively, one can also carry out measurement in momentum
space. The magnons associated with spin waves of wave number $k$
have momenta ${\bf p}=\hbar k {\bf e_y}$. If one observes the
Raman scattering using two Raman beams through the Bose gas, the
momentum conservation between the magnons and Raman photons
requires $\Delta {\bf k}=k {\bf e_y}$ with $\Delta {\bf k}$ being
the difference of wave vectors between two Raman beams. Hence the
momentum distribution of the scattered Raman photons can identify
the existence of the spin waves.

In conclusion, Bose condensates in an optical lattice offer a new
tool and test ground to study the quantum spin phenomena. The Bose
statistics in each lattice site makes the atoms coherently behave
as a spin magnet. Such an array of spin magnets not only exhibit
fascinating spin dynamics as shown in this letter, but also may
find potential applications in quantum information and
computation. This opens new opportunities for future research.

We would like to thank Dr. B. P. Anderson for several enlightening
discussions concerning the detection of magnons. This work is
supported in part by the US Office of Naval Research under
Contract No. 14-91-J1205, by the National Science Foundation under
Grant No. PHY98-01099, by the US Army Research Office, by NASA,
and by the Joint Services Optics Program.

\end{document}